\PassOptionsToPackage{dvipsnames,svgnames,x11names}{xcolor}
\documentclass[11pt]{article}
\usepackage[a4paper, total={6.3in, 9.7in}]{geometry}
\usepackage{setspace} 
\usepackage{authblk}
\usepackage{graphicx,xcolor}
\usepackage{doi}
\usepackage{cite}
\usepackage{booktabs}
\usepackage{amsmath,mathtools,amssymb}
\usepackage{hyperref}

\usepackage{etoolbox}

\makeatletter
\patchcmd{\env@cases}{1.2}{0.72}{}{}
\makeatother

\hypersetup{
    pdftitle={Exploration of Parameter Spaces Assisted by Machine Learning},
    colorlinks,
    linkcolor={red!50!black},
    citecolor={blue!50!black},
    urlcolor={blue!80!black}
}

\definecolor{LightGray}{gray}{0.9}

\title{\Large\bf Exploration of Parameter Spaces Assisted by Machine Learning}
\author[1]{A. Hammad\footnote{\href{mailto:ahhammad@cern.ch}{ahhammad@cern.ch}}}
\author[1,2,3]{Myeonghun Park\footnote{\href{mailto:parc.seoultech@seoultech.ac.kr}{parc.seoultech@seoultech.ac.kr}}}
\author[1]{Raymundo Ramos\footnote{\href{mailto:rayramosang@gmail.com}{rayramosang@gmail.com}}}
\author[1,2]{Pankaj Saha\footnote{\href{mailto:saha@seoultech.ac.kr}{saha@seoultech.ac.kr}}}
\affil[1]{\textit{Institute of Convergence Fundamental Studies, Seoultech, Seoul 01811, Korea}}
\affil[2]{\textit{School of Natural Sciences, Seoultech,  Seoul 01811, Korea}}
\affil[3]{\textit{School of Physics, KIAS, Seoul 02455, Korea}}
\date{}

\begin{document}

\maketitle

\begin{abstract}
\noindent We demonstrate two sampling procedures
assisted by machine learning models via regression and classification.
The main objective is the use of a neural network
to suggest points likely inside regions of interest,
reducing the number of evaluations of time consuming calculations.
We compare results from this approach with results from other sampling methods,
namely Markov chain Monte Carlo and MultiNest,
obtaining results that range from comparably similar to arguably better.
In particular, we augment our classifier method with a boosting technique
that rapidly increases the efficiency within a few iterations.
We show results from our methods applied to a toy model
and the type II 2HDM, using 3 and 7 free parameters, respectively.
The code used for this paper and instructions are publicly available on the web%
\footnote{\url{https://github.com/AHamamd150/MLscanner}}.
\end{abstract}

\newpage


\section{Introduction}
\label{sec:intro}

Technological advancements bring computers with more powerful processing capabilities
but at the same time bring more diverse, advanced and precise
experimental probes.
A considerable part of the scientific community
is tasked with applying these powerful computers
to adjust known and new theoretical models
to the most up to date constraints
found by experiments.
This usually involves taking a set of free parameters from the model,
calculate predictions for the observables that depend on them
and comparing with the experiments of interest.
Then we can judge the success of the model
based on how well it can predict said observables
within the experimental errors.
One common starting point for new models
is extending the model that works best, e.g.,
the standard model (SM) of particle physics for high energy physics (HEP)
or the cold dark matter model with a cosmological constant ($\Lambda$CDM) for cosmology.
For extensions that attempt to explain
many deviations observed in experiments
as well as provide missing pieces,
one may end up with a considerably large parameter space.
In the case of the SM,
for a supersymmetric extension
the full number of parameters reaches the order of hundreds~\cite{pdg2022},
and for experimental studies using the SM effective field theory (SMEFT)
the number of parameters could be several tens or more~\cite{Brivio:2017btx}.
Regarding the SMEFT, it has been show that machine learning techniques
can aid in the estimation of likelihood ratios and setting limits
in LHC analysis when large number of dimensions and many observables are considered~\cite{Brehmer:2018kdj,Brehmer:2018eca}.
With an increase of parameters the number of points required
for proper sampling increases exponentially---the well-known \textit{curse of dimensionality}.
Multiplying this number by the time required to calculate
an ever growing number of experimental constraints~\cite{pdg2022}
can give an estimation of the required effective time.

Besides parallelizing computations,
to simplify and accelerate the task of exploring large
parameter spaces (both in size and range of parameters),
several techniques efficiently use any information gathered about the problem
to infer parameter distributions.
Two successful and well known examples
are Markov chain Monte Carlo (MCMC)~\cite{Mackay:2003inf}
and MultiNest~\cite{Feroz:2007kg, Feroz:2008xx}
(see Ref.~\cite{Lewis:2002ah,Lafaye:2004cn,Bechtle:2004pc,RuizdeAustri:2006iwb,Allanach:2007qj,Strege:2014ija,Han:2016gvr,GAMBIT:2017yxo,Bagnaschi:2017tru,Brinckmann:2018cvx}
for examples of studies that use these tools).
These tools are cleverly designed
to use the results from likelihood calculations
to infer distributions of model parameters
and provide sampling consistent with this distribution.
Expectedly, in the course of working with these tools,
one may find particularly complicated subspaces
or regions with problematic features that may result
in excessive evaluations of the likelihood function
or poorly sampled regions~\cite{Ren:2017ymm,Goodsell:2022beo}.
It is worth noting that issues may be found for any sampling method
and depend on the implementation of the algorithm.

Machine learning (ML) techniques are natural contenders
for alleviating such difficulties
for their ability to find concealed patterns
in large and complex data sets
(See Ref.~\cite{Feickert:2021ajf} for a review and references therein).
In fact, in Ref.~\cite{Graff:2011gfh},
an algorithm for
blind accelerated multimodal Bayesian inference (BAMBI)
is proposed that combines MultiNest
with the universal approximating power of neural networks.
Recently, a general purpose method known as ML Scan (MLS)~\cite{Ren:2017ymm,Staub:2019xhl} was proposed,
where the authors used a deep neural network (DNN)
to iteratively probe the parameter space,
starting with points randomly distributed.
In there, the sampling is improved incrementally with active learning.
Conversely, the active learning methods proposed in Refs.~\cite{Goodsell:2022beo,Caron:2019xkx}
are based on finding \textit{decision boundaries} of the allowed subspace.
Application to HEP can be found in Ref.~\cite{Rocamonde:2022gyw}.
In Ref.~\cite{deSouza:2022uhk}, the authors introduce dynamic sampling techniques for beyond the SM searches.

In this work, we have implemented two broad classes of ML based efficient sampling methods of parameter spaces,
using regression and classification.
Similarly to Refs.~\cite{Ren:2017ymm,Staub:2019xhl,Goodsell:2022beo,Caron:2019xkx},
we employ an iterative process
where the ML model is trained on a set of parameter values
and their results from a costly calculation,
with the same ML model later used to predict
the location of additional relevant parameter values.
The ML model is refined in every step of the process,
therefore, increasing the accuracy
of the regions it suggests.
We attempt to develop a generic tool
that can take advantage of the improvements
brought by this iterative process.
Differently to Refs.~\cite{Goodsell:2022beo,Caron:2019xkx},
we set the goal in filling the regions of interest
such that in the end we provide a sample of parameter
values that densely spans the region as requested by the user.
Moreover, considering the steps followed in each iteration,
we consider ways to improve the efficiency,
particularly by employing a boosting technique
that improves convergence during the first steps.
With enough points sampled,
it should not be difficult to proceed with more detailed studies
on the implications of the calculated observables
and the validity of the model under question.
We pay special attention to give control to the user
over the many hyperparameters involved in the process,
such as the number of nodes, learning rate, training epochs, among many others,
while also suggesting defaults that work adequately in many cases.
The user has the freedom to decide whether to use regression or classification
to sample the parameter space,
depending mostly on the complexity of the problem.
For example, with complicated fast changing likelihoods
it may be easier for a ML classifier to converge
and suggest points that are likely inside the region of interest.
However, with observables that are relatively easy to adjust,
a ML regressor may provide information useful to locate points of interest
such as best fit points,
or to estimate the distribution of the parameters.
After several steps in the iterative process,
it is expected to have a ML model that can accurately predict
the relevance of a parameter point
much faster than passing through the complicated time consuming calculation
that was used during training.
As a caveat, considering that this process requires iteratively training a ML model
during several epochs, which also requires time by itself,
for cases where the calculations can be optimized to run rather fast,
other methods may actually provide good results in less time.

The organization of this paper is as follows. In Sec.~\ref{sec:mlapsf} we describe in detail
the two iterative processes for sampling parameter spaces
using regression and classification.
We expand in Sec.~\ref{sec:toymodels} by applying said processes
to two toy models in 2 and 3 dimensions,
including a description of how
we can boost the convergence of the ML model.
In Sec.~\ref{sec:mlthdm} we describe how we sample regions of interest
of the two Higgs doublet model (2HDM) using well-known HEP packages
and the processes described in this paper.
In the end, we conclude with a summary of the most relevant details presented in this work.

\section{Machine Learning Assisted Parameter Space Finding}
\label{sec:mlapsf}

The process starts with a set of random values, $K_0$,
for the parameters that will be scanned over,
and their results from a calculation of observables, $Y(K_0)$.
This random set of parameter values and their results
are used to train a ML model
with the purpose of predicting meaningful regions
in successive iterations where the model will be further refined.
After the initial ($0^\text{th}$) training step,
we will follow an iterative process
that, in its most basic form,
is summarized with the following steps:
\begin{enumerate}
    \item
        Use the ML model to obtain a prediction, $\hat{Y}$,
        for the output of a HEP calculation
        for a large set of random parameter values, $L$.
    \item
        Based on this prediction, select a smaller set of points, $K$,
        using some criteria that depends on the type of analysis of interest.
        Up to this step we should have a set $K$
        and its corresponding predictions $\hat{Y}(K)$.
    \item
        Pass the set $K$ of parameter values to the HEP calculation
        to obtain the actual results $Y(K)$.
    \item
        Use the set $K$ and its results $Y(K)$
        to refine the training of the ML model.
    \item
        The loop closes when we use this refined model
        to predict the HEP calculation output for a larger set of parameter values
        as in step 1 above.
\end{enumerate}
Later in the text and in specific examples
we will expand on the implementation of this brief summary.

\begin{figure}[tb]
    \center
    \includegraphics[width=\textwidth]{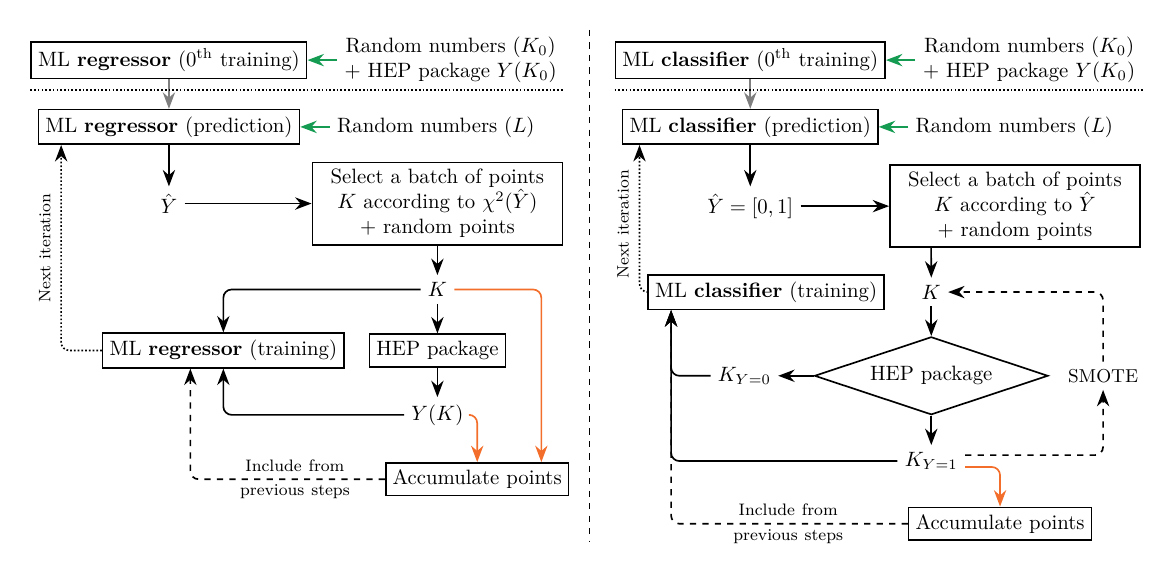}
    \includegraphics[width=0.8\textwidth]{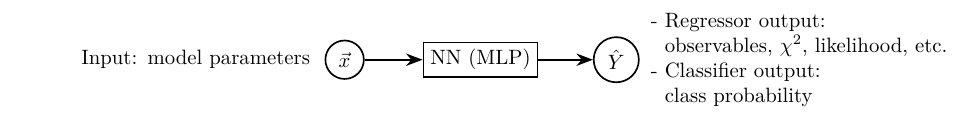}
    \caption{%
        Top: Charts for the ML iterative process
        used for the regressor (left)
        and the classifier (right).
        The main predict-train loop is indicated with black arrows,
        green arrows indicate places where a random number set is required
        and the orange arrow marks where we collect the output points.
        The points accumulated up to step $n_k$
        are added for training during step $n_{k+1}$.
        Bottom: The neural network used for the regressor and classifier
        will take model parameters as input.
        In the case of the regressor,
        returns observables, likelihood, or any other results
        that were used for training.
        For the classifier, it returns the class probability.
        The neural network we use
        in the examples of the text
        is a multilayer perceptron (MLP).
        }
    \label{fig:charts}
\end{figure}


It is assumed that the calculation of the observables
is done through an expensive, time consuming computation,
such that it is worthwhile spending additional time
training a ML model with its results.
Taking these basic steps as the starting point,
one can further fill in the details of the sampling,
considering that there is a number of ML models to choose from
and that the selection of the set $K$
depends highly on the region of interest.
Regarding the choice of the set $K$,
in general it is a good idea to add an amount of points chosen at random
to enforce exploration of new parameter space regardless of the suggestion of the ML model.
It is useful to gather the $K$ set and the results from the HEP calculation
in every iteration as they represent the sampling of the parameter space.
For the training step
we always use the new points from the corresponding iteration,
but we have the choice to add all or some of the set of accumulated points.
After a large amount of points have been accumulated
training on the full set may be time consuming,
therefore, it would be a good idea to devise rules on how to include them.
For example, in Ref.~\cite{Goodsell:2022beo} the full set of
accumulated points is used to train after a fixed number of iterations.
In what follows we will describe our implementation of two broad types of models:
ML regressors and ML classifiers.

\subsection{Sampling with a regressor}

We will use a ML regressor
whenever we are interested in training a model
to predict actual outputs from the calculation of observables.
In this case the training requires parameter values
and their numerical results from the HEP calculation.
When we pass a large number of parameter values
we get a set of predictions for the observables, $\hat{Y}$,
that can aid in the selection of the regions of interest.
In this case, the $K$ set could be composed of relevant points
based on, e.g., $\chi^2$ or likelihood values.
However, one could devise any number of selection mechanisms
that take advantage of the access to predictions for the observables.

One of the first complications of sampling with the regressor
is the possibility that the sample for training in each step
does not contain points close enough to the
regions of interest.
Another one is that succesive training sets may be heavily biased
towards particular results,
e.g., when most of the parameter space yields similar observations
and far from the ones we are interested in.
In the case of the first complication,
one may use known maximums as seeds and sample
around them.
For the second complication,
it is important to make a selection of points for training
that teaches the model the diversity of the output.
Otherwise we risk training a model that is biased
to predict results closer to the majority of the training set.
In fact, here is where the use of the predictions
from the model could have more power.
Using the model predictions we can infer a set of points
that could have a diverse and relevant set of results,
worthy of passing to the HEP calculation
and later to the training step.

The precision of the predictions is expected to improve with
more iterations and this can be checked easily via measures
such as the mean absolute error (MAE).
After several iterations, this process
should result in a parameter space sampling
heavily weighted towards the region of interest used to select the $K$ set
and a model specialized in giving accurate and fast predictions for this region.

\subsection{Sampling with a classifier}

A different approach would be training a ML classifier
to predict if a point in the parameter space
would be classified as inside or outside the region of interest
according to conditions set by the study being performed.
Examples of this conditions would be
whether or not the parameter point is theoretically viable
and if the point is still inside constraints set by experiments,
among several others options.
In contrast to the regressor described above,
in this case, we expect a HEP calculation with a simple binary output,
say, 1 if the point is inside a region of interest or 0 if otherwise.
After training the ML model,
the predictions, $\hat{Y}$, would be distributed in the range $[0,1]$.
This predictions can be considered as an encoding
of the confidence the model has to classify points in different regions.
This presents different opportunities for the selection of points.
For example, a simple choice would be to take $K$ from the points most likely to be inside.
Another choice could be taking points around $\hat{Y} = 0.5$,
where the model is more uncertain,
or even a combination of different ranges.

One advantage of having this classification of points,
is that we can try to balance training with sets of equal size for
the two classes of points and apply boosting techniques if any class is undersampled.
One example of boosting technique is the synthetic minority oversampling technique (SMOTE)~\cite{SMOTE}
that will be explained later in the text.
This is helpful in cases where one of the classes is difficult to sample randomly,
helping with the first few iterations when the model not accurate enough.
Another advantage of a classification is that it allows the accumulation
of a sample of points inside the allowed parameter space
in separation from the sample of points outside that needs to be kept
to train the model.

While for the regressor the objective was a model accurately trained
on highly relevant regions,
here the goal is to train a model that can accurately draw a boundary
between points of each class.
For the resulting samples, it is expected to see
a concentration of points around said boundary
indicating where the model has been trained more extensively.
Additionally, considering that the training
happens with balanced samples,
there could be a concentration of points
in the class with the smallest space,
which most of the times is the class of interest.
Obviously, the model can be used to predict
which additional points could belong to the
class we want to study.

\subsection{General comments}

It is entirely possible
that during training
the model specializes in the training set
and gives inaccurate predictions when
presented with new points.
However, in our approach this overtraining is avoided, in every iteration,
by comparing with the results from the true calculation and correcting regions where
the ML model is giving wrong advise.
For the regressor,
since we focused mostly on training with points in regions of large importance,
there is a chance of larger error in predictions
outside this region.
Obviously, the expectation is that the process outlined above
keeps this error under control to avoid missing relevant regions.
The case of the classifier is different.
Since here we deal with a calculation
that decides whether or not a point is in the target region,
the trained model responds with how confident
it is that a new point belongs inside or outside.
Therefore, we require the model to accumulate enough
confidence to give certain enough predictions for both classes.

In Fig.~\ref{fig:charts} we show flowcharts
of the iterative process followed
for both the ML regressor (left) and the ML classifier (right).
Both of them start with an initial $0^\text{th}$ training step
and proceed to the first prediction step inside the iterative steps.
The green arrows indicate the places
where sets of random parameters have to be inserted
and orange arrows indicate the steps where points are accumulated
in the sampled parameter space pool.

\section{Application to a multidimensional toy model}
\label{sec:toymodels}

In this section we apply the iterative process
described in the previous section
to a simple model in 3 dimensions.
For this model it is possible to find
several curved regions of interest that are completely disconnected
while also encircling regions of low relevance.
While these two features hardly capture all the complexity
that can appear when scanning the parameters of a realistic multidimensional model,
with this toy model it is shown how, by means of the processes outlined in the last section,
these two generic complications are easily addressed.

The 3-dimensional toy model is given by the function
\begin{equation}
    \label{eq:3dmodel}
    O_\text{3d} = \left[2 + \cos\left(\frac{x_1}{7}\right) \cos\left(\frac{x_2}{7}\right) \cos\left(\frac{x_3}{7}\right)\right]^5
\end{equation}
where we will use a preferred region
with center at $c_\text{3d} = 100$
and a standard deviation of $\sigma_\text{3d} = 20$.
We will take a gaussian likelihood given by
$\mathcal{L}_\text{3d} = \exp[-(O_\text{3d} - c_\text{3d})^2/2\sigma_\text{3d}^2]$.
This choice of central values and deviations
results in shell-shaped regions of interest
for this 3-dimensional model.
For all the dimensions of the toy model ($j=1,2,3$)
we will scan in the range $x_j \in [-10\pi,10\pi]$.

\begin{table}[tb]
    \setlength\tabcolsep{0.2cm}
    \begin{tabular}{lcc}
        \toprule
         & Regressor & Classifier \\
        \midrule
        size of initial set & 500 points & 500 points \\
        size of test set ($L$) & 50\,000 points &  50\,000 points \\
        size of selection each step ($K$) & 500 points & 500 points \\
        random points in $K$ & 10\% & 10\% \\
        input layer (IL) dimension & 3 & 3 \\
        hidden layers (HL) & 4 & 4 \\
        HL neurons & 100 & 100 \\
        HL activation function & ReLU & ReLU \\
        output layer (OL) dimension & 1 & 1 \\
        OL activation function & \textbf{linear} & \textbf{sigmoid} \\
        loss function & \textbf{mean absolute error} & \textbf{binary cross entropy} \\
        optimizer & Adam ($\beta_1 = 0.9$, $\beta_2 = 0.99$) & Adam ($\beta_1 = 0.9$, $\beta_2 = 0.99$) \\
        learning rate & 0.001 & 0.001 \\
        epochs & 1000 & 1000 \\
        \bottomrule
    \end{tabular}
    \caption{\label{tab:toymodels}%
        Hyperparameters used for the results displayed in Fig.~\ref{fig:3dregclass}
        for the toy model.
        The parts that differ between regressor and classifier are shown in boldface text.
        }
\end{table}

To the toy model we will apply
sampling with both regression and classification
to illustrate the advantages of each approach
and demonstrate the differences in the obtained results.
The particular type of DNN
that we will use for these examples is a multilayer perceptron (MLP)
with fully connected layers.
The summary of hyperparameters that we use
is contained in Table~\ref{tab:toymodels}.

\subsection{Exploring the toy model with the regressor}

The neural network for the regressor
will be constructed with 4 hidden layers,
each of them with 100 neurons,
and rectified linear unit (ReLU) activation function.
The final layer will have one neuron
with a linear activation function.
For the loss function we will use the mean absolute error.
To optimize the weights we will use the Adam algorithm with a learning rate of 0.001,
exponential decay rates $\beta_1 = 0.9$ and $\beta_2 = 0.999$,
and train during 1000 epochs.

To test the ML regressor,
in every iteration
we make predictions for the observable from a large set of points, $L$,
that we use to calculate the likelihood.
Then we select a smaller set of points, $K$,
A fraction of 90\% of the set $K$ will be selected with a likelihood above 0.9,
while the other 10\% will be taken randomly from the parameter space.
Remember that this set $K$
contains the points
for which we will calculate
the actual observable and likelihood.
In every iteration we accumulate all the $K$ sets and their results
and this accumulated set constitutes the output sample and also the training set.
For training, we will include all the accumulated points
every 2 iterations and in the final iteration
to improve the accuracy of the model,
in other iterations we use a partial set.
We accumulate all the suggested points
which after a few iterations should start to accrue around the region of interest
($\mathcal{L}_\text{3d} > 0.9$).

For the sample displayed in the upper left panel of Fig.~\ref{fig:3dregclass}, under DNNR,
we stop the process after collecting 20\,000 points with the condition
$\mathcal{L}_\text{3d} > 0.9$.
we start with an initial set of 500 random points
and in every iteration we use the DNN to predict the value
of 50\,000 points and select 500
that are passed to the likelihood calculation.
It is worth noting in this point,
that if the likelihood had to be computed
via a complicated, heavy, time consuming calculation
we would be using the DNNR to guess the values
of 100 times more points than we pass to the actual calculation.
Here is where the process described above could be more powerful.

\subsection{Exploring the toy model with the classifier}

The construction of the neural network for the classifier
is very similar to the construction used for the regressor.
We again use 4 hidden layers with 100 neurons each and ReLU
activation function.
The output layer has an output of 1 dimension
and we use the sigmoid activation function
to obtain values between 0 and 1.
the output of the classifier
consist on the probability
that the input belongs in the class.
In this case the class will be defined as points with
$\mathcal{L}_\text{3d} > 0.9$.
For the loss function we use binary cross entropy
and we train with the Adam optimizer
using a learning rate of 0.001,
and parameters $\beta_1 = 0.9$ and $\beta_2 = 0.999$.
The training happens during 1000 epochs.

In the case of the classifier,
the output of the DNN is a probability
of how confident the model is that the point
belongs to the class.
Considering this,
we base our selection of points on this confidence, $\hat{Y} \in [0, 1]$,
calculated from the larger set $L$.
The smaller set $K$ will be made with a fraction of 10\% points chosen randomly
as with the regressor.
For the remaining 90\%, we will choose 50\%
from points such that $\hat{Y} > 0.75$,
and 50\% from points predicted uncertain or outside, $\hat{Y} < 0.75$.
There is nothing special about the number 0.75,
but it is convenient to take a number
that divides the set of points
in one with points likely inside
and another with points that are uncertain or unlikely to be
in the class.
As before, this set $K$ is passed to the calculation
that outputs the correct classification.
Differently from the regressor,
in this case we can accumulate only the points that belong to $\mathcal{L}_\text{3d} > 0.9$
while the points that are outside the class can be kept
for training.
If after obtaining the correct classes
the training set is dominated by points inside or outside the class,
we need to perform an extra step
where the minority class is oversampled
or the majority class undersampled.
This is to ensure that the network learns
the distribution of both sets properly.
To oversample the minority class
we can use a boosting method
as will be explained after the next subsection.
Something to note here
is that it is important to train the network
on the points that where missclassified
and the points where $\hat{Y} \approx 0.5$.

\begin{figure}[tbh!]
    \centering
    \includegraphics[scale=0.9]{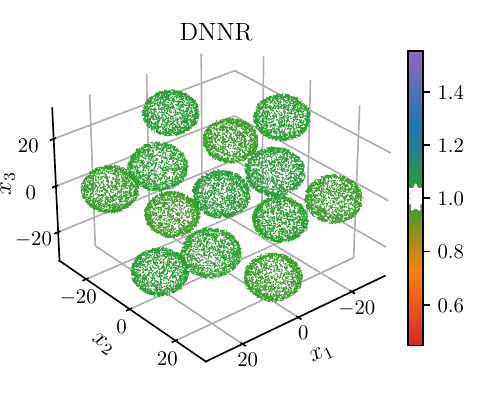}%
    \includegraphics[scale=0.9]{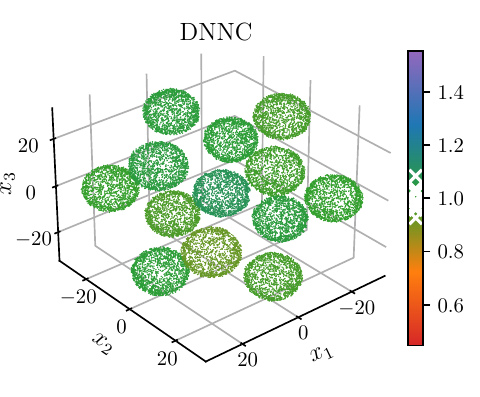}
    \includegraphics[scale=0.9]{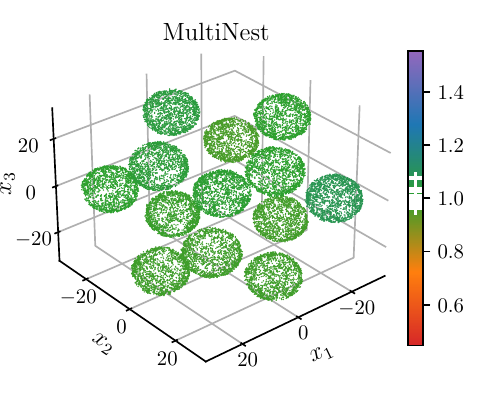}%
    \includegraphics[scale=0.9]{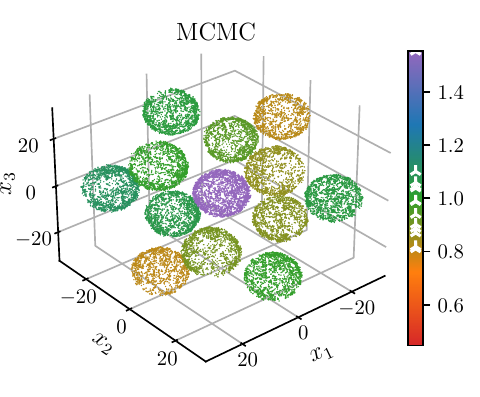}
    \includegraphics[scale=0.8]{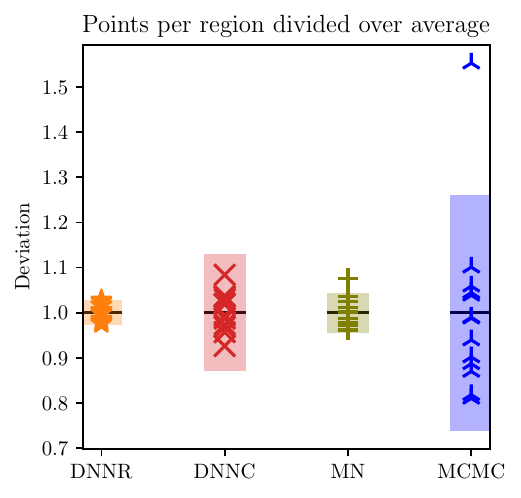}
    \caption{%
        Coverage for 13 different regions of the toy model with 20\,000 points,
        using DNNR (top left), DNNC (top right), MultiNest (middle left)
        and MCMC (middle right).
        The color indicates deviation from average distribution of points per region
        (with average normalized to 1).
        Best result (less deviation) out of 10 runs is displayed.
        In the bottom panel, in light color, we show standard deviations for the 13 regions averaged over 10 runs.
        Again, the average number of points per region has been normalized to 1.
        The markers show the distribution of the best result out of 10, corresponding to the points in the top 4 panels.
    }
    \label{fig:3dregclass}
\end{figure}

The sample with the most uniform distribution out of 10 runs
is displayed in the upper right panel of Fig.~\ref{fig:3dregclass}.
For this figure we collected 20\,000 points that belong to the class
$\mathcal{L}_\text{3d} > 0.9$.
In this case we attempt to train with an initial set of 500 random,
and in every iteration obtain predictions for 50\,000 test points (set $L$).
We select 500 points (set $K$) according to the rules stated above
and calculate the true classes.
Due to the balancing of the classes, in the first few steps
the algorithm may oversample or undersample imbalanced sets
and change the size of the set $K$.

\subsection{Comparison of coverange with other sampling methods}

To compare the coverage against other sampling methods,
we use \texttt{PyMultiNest}~\cite{Buchner:2014nha} and \texttt{emcee}~\cite{Foreman:emcee} implementations
of MultiNest~\cite{Feroz:2007kg,Feroz:2008xx} and MCMC~\cite{Mackay:2003inf}, respectively.
For \texttt{emcee} we use 500 walkers with other setting left as default.
In the case of \texttt{PyMultiNest} we increased the number of active points
to 6200 (\texttt{n\_live\_points}) and reduced the efficiency to 0.5 (\texttt{sampling\_efficiency}).
The increase of live points is aimed at collecting around 20\,000 points with $\mathcal{L}_\text{3d} > 0.9$.
In order to force both \texttt{emcee} and \texttt{PyMultiNest}
to concentrate on the region with $\mathcal{L}_\text{3d} > 0.9$,
we apply a weight depending on the value of $\mathcal{L}_\text{3d}$ according to
\begin{equation}
    \mathcal{W} =
        \begin{cases}
            \epsilon & \text{if } \mathcal{L}_\text{3d} \leq 0.89 \\
            1 & \text{if } \mathcal{L}_\text{3d} \geq 0.91 \\
            50 (1 - \epsilon)(\mathcal{L}_\text{3d} - 0.89) + \epsilon & \text{if } 0.89 < \mathcal{L}_\text{3d} < 0.91
        \end{cases}
\end{equation}
where the last case is just $\mathcal{W}$ growing linearly to connect the two different $\mathcal{L}_\text{3d}$ regimes.
For all the examples shown in Fig.~\ref{fig:3dregclass},
we run the sampling method 10 times,
storing around 20\,000 points with the condition $\mathcal{L}_\text{3d} > 0.9$.
In the four upper panels,
we show the samples accumulated for each method
with the most consistent distribution of points
among disconnected regions.
The color represents deviation from the average number of points in each region,
with the average normalived to 1.
Expectedly, using MCMC results in the least uniformly covered regions (most color variation)
with DNNR having the most consistent coverage (least color variation).
Visually, the spread of points using DNNR and DNNC appears more
uniform that in the other two examples.
In the bottom panel of the same figure
we show the average deviation for all the 10 runs
in solid light color.
The strong colored markers
show the deviations in all 13 regions
for the samples with the least deviation,
corresponding to the 4 upper panels.
Again, the deviations correspond to the number of points per region
with average normalized to one.
The DNNR shows the least deviation ($\pm 0.03$)
while DNNC sits at $\pm 0.13$.
DNNC performs somewhere in between MultiNest ($\pm 0.044$)
and MCMC ($\pm 0.26$).
The red $\times$ markers for DNNC
show that it is possible to have runs
with much less deviation than the average.

To comment on the most notable difference between the ML regressor and classifier,
the fact that the classifier explicitly distinguishes two classes
makes possible to operate differently on the points
depending on where they have been classified.
Above we mention that we have control
over how we pick up points for the training step
from the accumulated pools of the two classes.
To expand on the classifier,
in the next section we show
how we can use a small amount of points
to actively balance situations where
one of the classes tends to be oversampled.


\subsection{Toy model in higher dimensions}
\label{sec:toyNdimensions}

\begin{figure}
    \centering
    \includegraphics[scale=0.9]{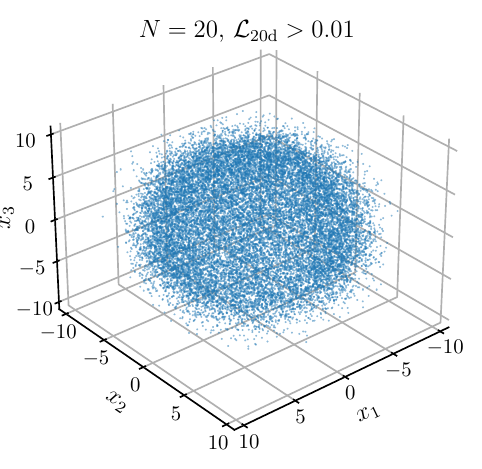}
    \includegraphics[scale=0.9]{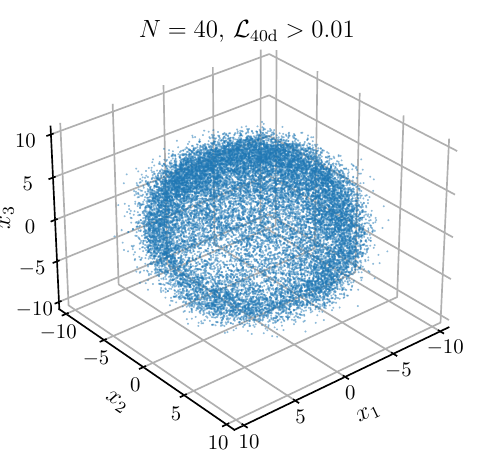}
    \caption{\label{fig:Ndregressor}%
        Result of accumulating 20\,000 points with $\mathcal{L}_{N\text{d}} > 0.01$
        for the toy model described in Sec.~\ref{sec:toyNdimensions}
        using the regressor.
        Only the first three dimension are shown that are shared with the 3-dimensional
        toy model of Eq.~\eqref{eq:3dmodel}.
        The model explored in the left panel has 20 dimensions
        while for the right panel it has 40 dimensions.
        A slight transparency has been applied to show
        parts with deficient coverage.
        }
\end{figure}

To display convergence in higher dimensions,
we apply the idea above to an $N$ dimensional toy model by extending
the toy model of Eq.~\eqref{eq:3dmodel}
with an extra toy observable in $N-3$ dimensions
\begin{equation}
    \label{eq:Nm3dmodel}
    O_{N-3\text{d}} = \sqrt{\sum_{j=4}^N x_j^2 }\,,
\end{equation}
which can be interpreted as a radius using the last $N-3$ dimensions
of the $N$ dimensional toy model.
We will require $O_{N-3\text{d}}$ to satisfy a central value $c_{N-3\text{d}}$
with corresponding error $\sigma_{N-3\text{d}}$.
The total likelihood will be given by
\begin{equation}
    \label{eq:Ndlikelihood}
    \mathcal{L}_{N\text{d}} = \exp\left[
        - \frac{(O_\text{3d} - c_\text{3d})^2}{2\sigma_\text{3d}^2}
        - \frac{(O_{N-3\text{d}} - c_{N-3\text{d}})^2}{2\sigma_{N-3\text{d}}^2}
    \right]\,.
\end{equation}
This can be seen as trying to find the regions around $O_\text{3d} \sim c_\text{3d}$ in three dimensions
while the other $N-3$ dimensions try to accommodate such that $O_{N-3\text{d}} \sim c_{N-3\text{d}}$.
We test the process described above with $N = 20$ and $N = 40$.
For $c_\text{3d}$ and $\sigma_\text{3d}$, as before, we use 100 and 20, respectively.
For the $N-3$ dimensional part we use $c_{N-3\text{d}} = 0.5$ and $\sigma_{N-3\text{d}} = 0.2$.
For the first 3 dimensions we scan in the range $[-10\pi/3, 10 \pi/3]$,
and for the rest of the dimensions we scan in the range $[-1, 1]$.
For this two cases,
the biggest change is the need to concentrate around regions of larger likelihood
when suggesting the large set $L$.
Therefore, in every iteration the set $L$ contains points
closer to points with large likelihood
found until that iteration.
This points are selected using gaussian jumps
with standard deviation that varies between 1/10 to 1/3 of the dimensions ranges.
Moreover, training in higher dimensions, expectedly, requires more points,
therefore, we decrease the threshold value
of the points we want to accumulate
to $\mathcal{L}_{N\text{d}} > 0.01$.
We also increased the number of points in the large set $L$ to $10^7$
and in the selected set $K$ to 5000.
The result from accumulating 20\,000 points for $N=20$
and $N=40$ using the regressor is shown in Fig.~\ref{fig:Ndregressor}.
The coverage is less uniform than in the 3-dimensional case
shown in Fig.~\ref{fig:3dregclass}.
Particularly, in the 40-dimensional case,
some parts show a more sparse coverage.
Remember that the inside of the ball-like regions of interest
have a low likelihood region,
so it is not expected that the center is densely filled.
The results using the classifier are similar and, therefore, not displayed.
Considering that the points tend to appear distributed and not excessively clumped,
the sparsity could be reduced by a longer running time and/or removing a burn-in set.

\subsection{Boosting the start up convergence}
\label{sec:3.1}

Unlike the ML regressor, in the case of the ML classifier
both classes of points are of equal importance.
Accordingly, points outside the region of interest are also accumulated
to train the model to find their subspace.
Therefore, we train the model using data sets of the same size.
When the subspace of points of one class
is comparably much smaller than that of the other class,
the ML model is inefficient to suggest enough points of the smaller class
to pass to the HEP calculation in the initial steps,
leading to slow convergence.
Moreover, we end up training the model with much more points from one class than the other.
There are several ways to rectify this imbalance of the data:
\begin{itemize}
    \item Undersampling the majority class. In this case we loose much of the information from the undersampled class causing the model to converge more slowly.
    \item Oversampling the minority class by creating random copies from the underrepresented set~\cite{Goodsell:2022beo}. This makes the model overestimate the majority class during the test step.
    \item Oversampling the minority class using SMOTE~\cite{SMOTE}. The SMOTE transformation is an oversampling technique in which the minority class is enhanced by creating synthetic samples rather than by oversampling with replacement.
        This technique involves taking each point from the minority class and introducing synthetic samples along the line segments joining them to their nearest neighbors. Depending on the amount of oversampling required, a number $n$ of nearest neighbors can be randomly chosen.
        After this, a synthetic sample is generated
        from the lines between a point and its $n$ nearest neighbors
        depending on the amount of oversampling required.
 
\end{itemize}

\begin{figure}[tbh!]
    \centering
    \includegraphics[scale=0.85]{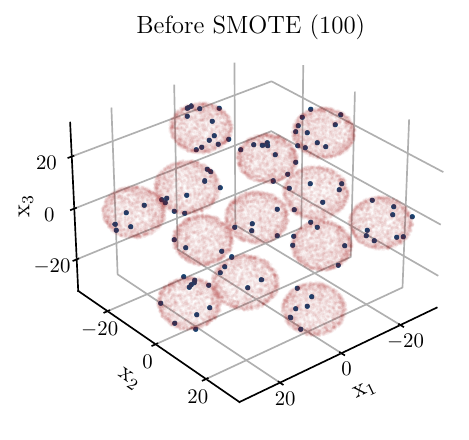}%
    \includegraphics[scale=0.85]{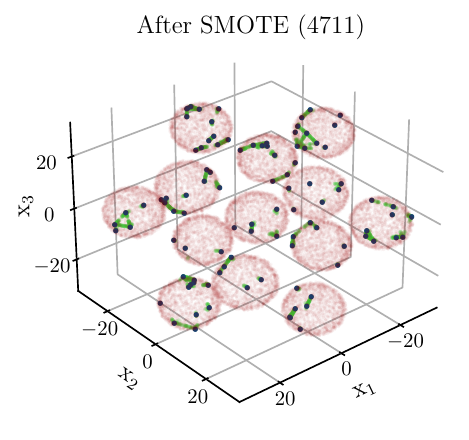}
    \includegraphics[scale=0.85]{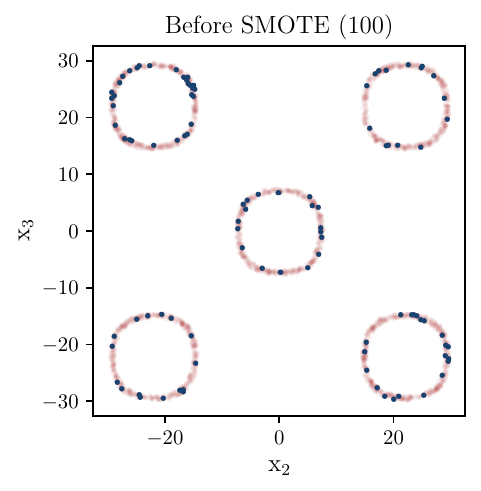}%
    \includegraphics[scale=0.85]{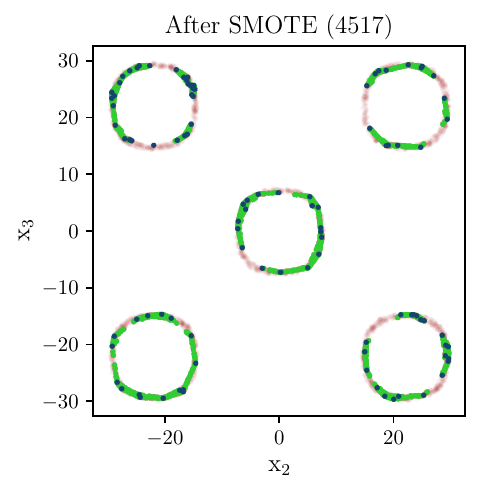}
    \caption{%
        Effect of the SMOTE technique on the toy model of Eq.~\eqref{eq:3dmodel}.
        On the upper panels,
        an initial set of 100 points (left, dark blue) in the target region
        is increased to 4743 (right, dark blue and green).
        In the bottom panels, we apply SMOTE to the $x_1 = 0$ slice of Eq.~\eqref{eq:3dmodel}
        to increase 100 points in the target region to 4517, where colors have identical meaning to the upper panels.
        In both upper and bottom panels the SMOTE technique boosts the minority class using 3 nearest neighbors.
        The proportion of points suggested by SMOTE that are actually
        in the target region is around 50\% for the upper panels
        and around 90\% for the bottom panels.
        The light red points indicate the target region in all the panels.
        }
    \label{fig:SMOTE1}
\end{figure}

The result of oversampling the minority class using SMOTE is shown in Fig.~\ref{fig:SMOTE1} where the minority class (dark blue points) with 100 points is oversampled to around 5000 points according to the description of the process given above.

\begin{figure}[tbh!]
    \centering
    \includegraphics[scale=0.5]{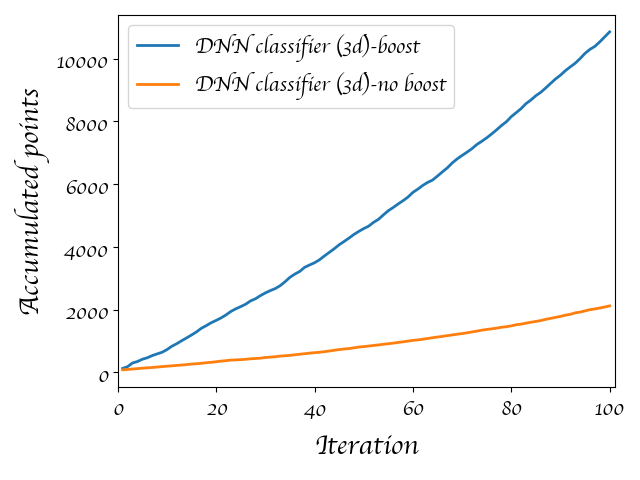}
    \caption{%
        Number of accumulated points in terms of iterations when using (blue) and not using (orange) the boosting method, for the 3-dimensional toy model.
       The sizes of the sets $K_0$, $L$ and $K$ as well as details for the selection of points are described in the text.}
    \label{fig:boost}
\end{figure}

As mentioned above, we attempt to take a set $K$ made of 90\% with predicted $\hat{Y} > 0.5$
expecting that the HEP calculation finds most of them with $Y=1$, while the other 10\% is taken randomly.
If the corrected classes, as given by the HEP calculation, have different sizes we use SMOTE to oversample the minority class.
The points suggested by SMOTE are then passed to the HEP calculation for proper classification.
It is important to mention that the number of the nearest neighbors used by SMOTE is a hyperparameter that has to be adjusted according to the study under consideration.
In Fig.~\ref{fig:boost} we compare the number of accumulated points after 100 iterations when SMOTE is applied (orange) and not applied (blue) to oversample the minority class ($Y = 1$ in this case) and in case of undersampling the majority class.
We found that for the 3-dimensional toy model, Eq.~\eqref{eq:3dmodel}, with $K_0=10\,000$, $L=3000$ and $K=200$
the model is able to accumulate 5 times more points when SMOTE is employed.
 
\section{Learning the Higgs signal strength in type II 2HDM}
\label{sec:mlthdm}

In this section we show the performance of ML models scanning over the
parameter space of the type II 2HDM scalar potential to match the measured
Higgs signal strength.

\subsection{The model}
\label{sec:themodel}

The 2HDMs~\cite{Lee:1973iz,Branco:2011iw} are extensions of the SM scalar sector containing two $SU(2)_L$ doublets, $\phi_1$ and $\phi_2$, sharing identical charge assignments under the SM gauge symmetry group:
\begin{equation}
	\phi_1 = \begin{pmatrix}
	\eta_1^+ \\
	(v_1 + h_1 + i h_3)/\sqrt{2} \\
	\end{pmatrix},\qquad
	\phi_2 = \begin{pmatrix}
	\eta_2^+ \\
	(v_2 + h_2 + i h_4)/\sqrt{2} \\
	\end{pmatrix}\,.
\end{equation}
The 4 real fields are labeled $h_i$, with $i=1,\ldots,4$;
the complex charged fields are $\eta_i^+$, with $i=1,2$;
the vacuum expectations values (VEVs) are $v_i$, with $i=1,2$.

The scalar setup in the 2HDM allows for flavor changing neutral currents (FCNC) from the Yukawa terms, which are restricted by current measurements.
A 2HDM with no FCNC can be obtained by adding a softly broken global $Z_2$ symmetry~\cite{Glashow:1976nt}
where $(\phi_1, \phi_2) \to (\phi_1, -\phi_2)$. In this case, the most general scalar potential is given by
\begin{equation}
\begin{split}
    V_\phi = & m^2_{11} (\phi^\dagger_1 \phi_1) + m^2_{22} (\phi^\dagger_2 \phi_2) - \left[m^2_{12}(\phi^\dagger_1\phi_2)+\text{h.c.}\right]+ \lambda_1 (\phi^\dagger_1 \phi_1)^2+ \lambda_2 (\phi^\dagger_2 \phi_2)^2  \\
    & +\lambda_3 (\phi^\dagger_1 \phi_1) (\phi^\dagger_2 \phi_2) +\lambda_4 (\phi^\dagger_1 \phi_2) (\phi^\dagger_2 \phi_1)  +\frac{1}{2}\left[ \lambda_5 (\phi^\dagger_1\phi_2)^2+\text{H.c.}\right]  \,,
\end{split}
\label{eq:Vphi}
\end{equation}

where $m_{12}^2$ softly breaks the $Z_2$ symmetry.
In $V_\phi$, the parameters $m_{jj}^2$ and $\lambda_{i\neq 5}$ are real, while $ m^2_{12}= |m^2_{12}|e^{i\eta(m^2_{12})}$ and $\lambda_5= |\lambda_5|e^{i\eta(\lambda_5)}$ are complex parameters that allow for CP violation~\cite{Antusch:2020ngh,Antusch:2021oit}.
In the following we consider a real potential with vanishing complex phases,  $\eta(\lambda_5)=\eta(m^2_{12}) =0$.
The VEVs $v_1$ and $v_2$ are related to the SM VEV by $v=\sqrt{v_1^2 + v_2^2} \sim 246$~GeV,
 and their ratio is defined as $\tan\beta \equiv v_2/v_1$.

The mass terms, $m^2_{11}$ and $m^2_{22}$, are determined from the minimization conditions of the scalar potential.
Four physical scalars are obtained after diagonalizing the mass matrices, with two CP even Higgses ($h_1, h_2$), one CP odd scalar ($A$) and one charged Higgs ($H^\pm$) with masses given by
\begin{align}
    \label{eq:massh1h2}
    m^2_{h_{1,2}} & = \frac{1}{2}\left[\chi^2_{11}+\chi^2_{22}\mp \sqrt{(\chi^2_{11}-\chi^2_{22})^2+4(\chi^2_{12})^2}   \right]\,, \\
    \label{eq:masshA}
    m^2_A  &= \frac{2m^2_{12}}{\sin2\beta}-\lambda_5 v^2\,,\\
    \label{eq:massHpm}
    m^2_{H^\pm}  &= \frac{2m^2_{12}}{\sin2\beta}-\frac{1}{2}(\lambda_4+\lambda_5) v^2\,,
\end{align}
with
\begin{align}
    \label{eq:mass11}
    \chi^2_{11} & =m^2_{12}\tan\beta+2\lambda_1 v^2\cos^2\beta\,,\\
    \label{eq:mass22}
    \chi^2_{22} & =m^2_{12}\cot\beta+2\lambda_2 v^2\sin^2\beta\,,\\
    \label{eq:mass12}
    \chi^2_{12} & =-m^2_{12}+\frac{1}{2}(\lambda_3+\lambda_4+\lambda_5) v^2\sin2\beta\,.
\end{align}

For the type-II 2HDM, the Yukawa terms that respect the $Z_2$ symmetry are written in the form
\begin{equation}
    -\mathcal{L}_Y = Y_u \overline{Q}_L\tilde{\phi}_2 u_R + Y_d\overline{Q}_L\phi_1 d_R + Y_\ell\overline{L}_L\phi_2 \ell_R + H.c\,, 
\end{equation}
where $\tilde{\phi}_2 = i\tau_2\phi_2^*$, $\tau_2$ is the $SU(2)$ generator corresponding to the Pauli matrix $\sigma_2$ and $Y_u,Y_d,Y_\ell$ are $3\times 3$ Yukawa coupling matrices.

We perform a scan over seven free parameters of the potential, $\lambda_j$ ($j\in \{1,2,3,4,5\}$), $\tan\beta$ and soft $Z_2$-breaking mass $m^2_{12}$,
to adjust the SM-like Higgs properties to match current measurements, taking into account other constraints from the electroweak global fit and B meson decays.
Note that one has the freedom to scan over physical parameters instead.
Each parameter base has their own advantages, for example,
using parameters of the potential allows to choose ranges
where stability and perturbativity test are automatically passed.
We use SPheno-4.0.5~\cite{Porod:2011nf} to calculate the particle spectrum of the physical eigenstates, \texttt{FlavorKit}~\cite{Porod:2014xia} for B meson decays while \texttt{HiggsBounds}-5.3.2~\cite{Bechtle:2008jh,Bechtle:2011sb} and \texttt{HiggsSignals}-2.2.3~\cite{Bechtle:2013xfa,Stal:2013hwa} are used to constraint the parameter space using recent Higgs boson measurements.
\texttt{HiggsBounds} constrains the parameter space by computing the theoretical prediction for the most sensitive channel for each scalar, $h_i$, and dividing by the observed experimental value to obtain the ratio, $\mathcal{O}_i$.
The computation of $\mathcal{O}_i$ requires production cross sections, $\sigma$, and decay branching ratios, $\mathrm{BR}(h_i)$,
as in
\begin{equation}
    \mathcal{O}_i = \frac{[\sigma\times \mathrm{BR}(h_i)]_{\text{model}}}{[\sigma\times \mathrm{BR}(h_i)]_{\text{obs}}}\,,
\end{equation}

where a value $\mathcal{O}_i > 1$ corresponds to a parameter point excluded by the $95\%$~C.L. limit.
\texttt{HiggsSignals} constrains the parameter space by evaluating the statistical compatibility of the SM-like Higgs boson in the model using recent data for the $125$ GeV Higgs resonance observed at the LHC\@.
\texttt{HiggsSignals} reports a total $\chi^2$ value for testing the model hypothesis as a combination of $\chi^2_\mu$ from the signal strength modifiers and $\chi^2_{m_{h_i}}$ from corresponding predicted Higgs masses as
\begin{equation}
    \chi^2_\text{tot} = \chi^2_\mu + \sum^{N_h}_{i=1} \chi^2_{m_{h_i}}.
\end{equation}
The best fit value is calculated according to
\begin{equation}
    \chi^2_\text{best} = \chi^2_\text{min}/n_\text{D.O.F.}\,.
\end{equation}
with $\chi^2_\text{min}$ the minimum $\chi^2$.
We adjust our selection to accept all points with $\chi_\text{tot}^2\le 95$, with a SM Higgs mass uncertainty of $\pm 2$~GeV. It is worth mentioning that all selected points are required to pass the \texttt{HiggsBounds} selection.

\subsection{Additional constraints}
\label{sec:addconst}

\begin{figure}[tbh!]
    \centering
    \includegraphics[scale=0.42]{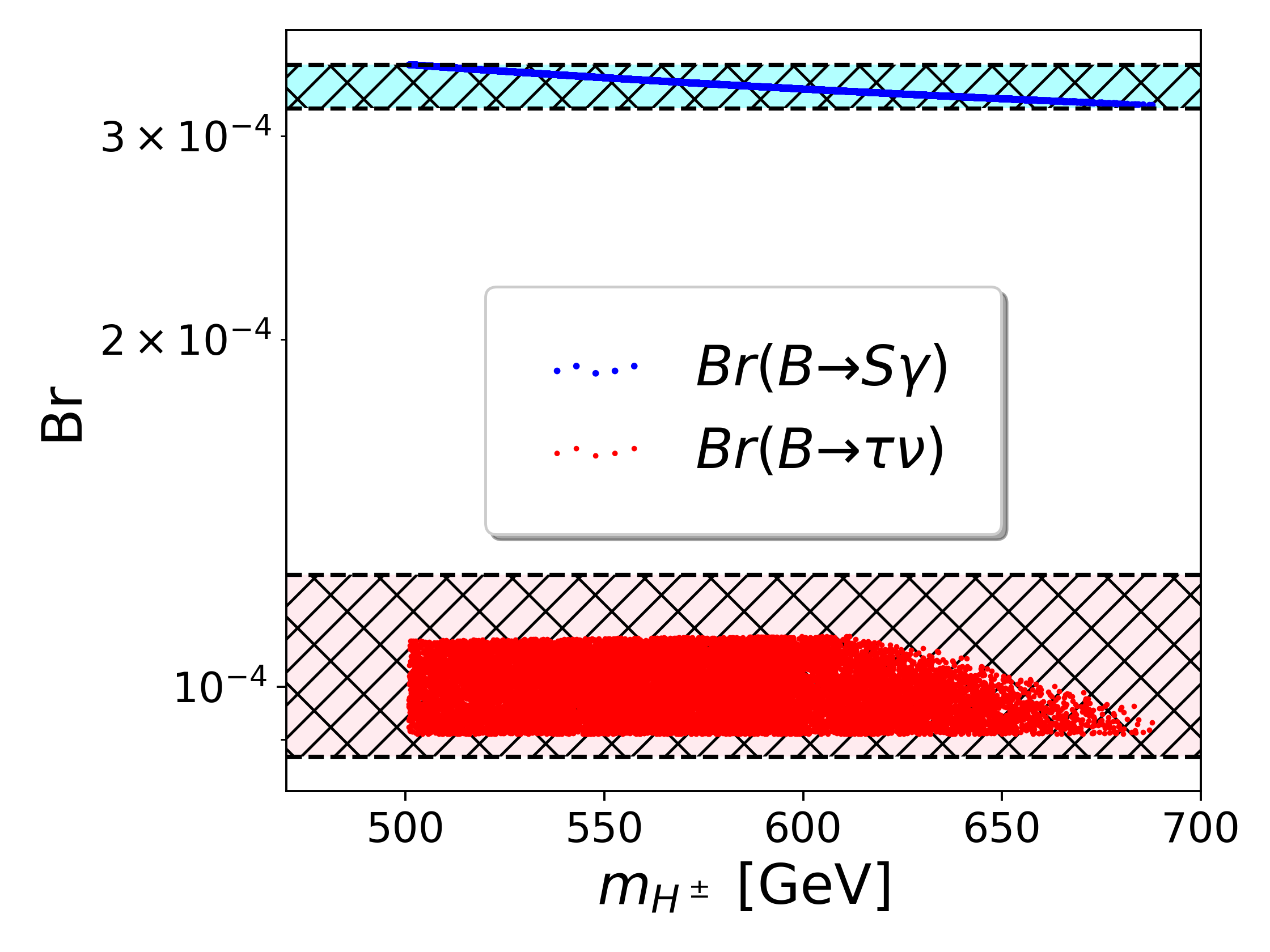}\hspace{0.7cm}
    \includegraphics[scale=0.4]{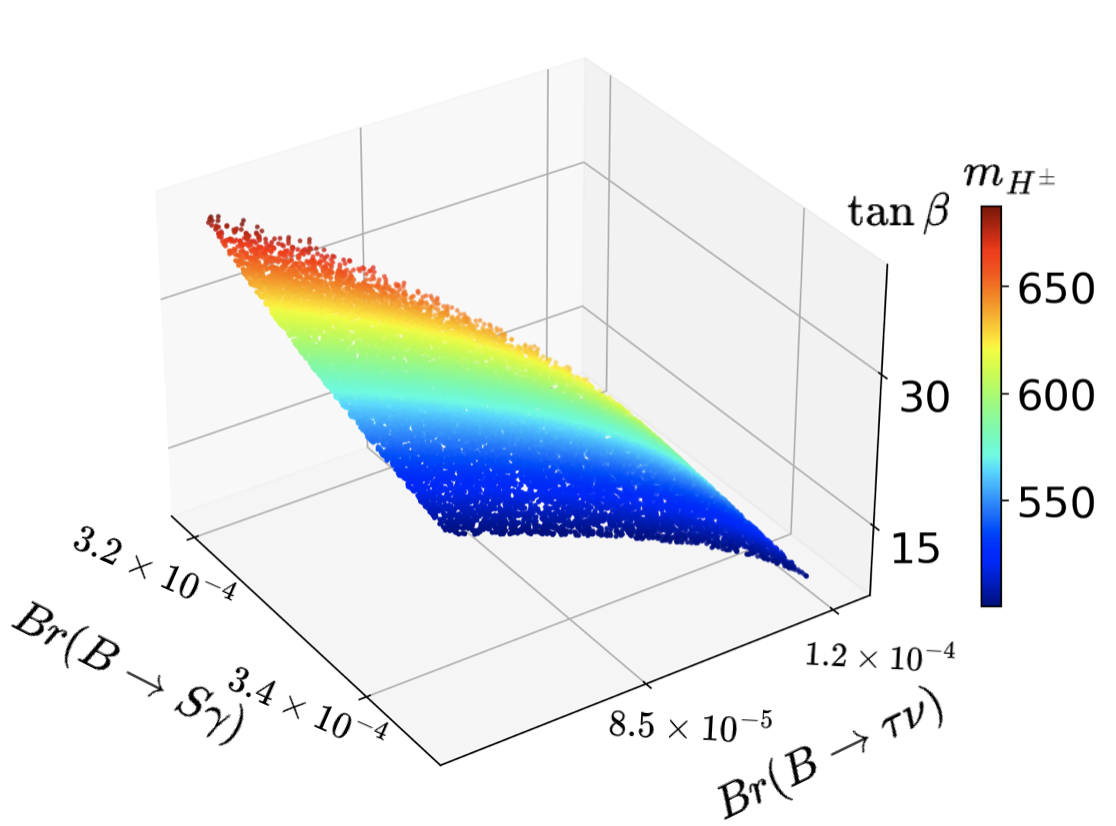}
    \caption{Accumulated points from  MLP  classifier model. Left: branching ratios of $B\to S\gamma$ (blue points) and $B\to\tau\nu$ (red points) versus the charged Higgs mass, while the shaded cyan and pink areas are the current allowed regions from $B\to S\gamma$ and $B\to\tau\nu$ respectively at re-fitted $1 \sigma$ reported in~\cite{HFLAV:2019otj}.
    Right:  Relationship between both branching ratios displayed on the left and $\tan\beta$, colored according to the value of the charged Higgs mass.
    }
\label{fig:B_constraints}
\end{figure}

The oblique parameters from the electroweak observables fit receive contributions from the 2HDM at one-loop level,
constraining the 2HDM parameter space.
From the global fit of the oblique parameters we have~\cite{Baak:2012kk}:
\begin{equation}
S = 0.03\pm0.10\,, \hspace{4mm} T = 0.05\pm0.12\,,\hspace{4mm} U = 0.03\pm0.10\,.
\end{equation}
Additionally, measurements from B meson decays add extra constraints on $(m_{H^\pm}\text{-}\tan\beta)$ plane.
For large $\tan\beta$,
the dominant constraints come from $\mathrm{BR}(B^+\to\tau^+\nu) = (1.06 \pm 0.19)\times 10^{-4}$ 
and $\mathrm{BR}(B\to S\gamma)_{E_\gamma \ge 1.6 \text{GeV}} = (3.32\pm 0.15)\times 10^{-4}$\cite{HFLAV:2019otj}.
The ranges for mentioned branching ratios are displayed in the left side of Fig.~\ref{fig:B_constraints}.

\subsection{Numerical scan details}

\begin{table}[tb]
    \setlength\tabcolsep{0.2cm}
    \begin{tabular}{lcc}
        \toprule
         & Regressor & Classifier \\
        \midrule
        size of initial set & 10, 1000 points & 10, 1000 points \\
        size of test set ($L$) & 200\,000 points &  200\,000 points \\
        size of selection each step ($K$) & 300 points & 300 points \\
        random points in $K$ & 20\% & 20\% \\
        input layer (IL) dimension & 7 & 7 \\
        hidden layers (HL) & 4 & 4 \\
        HL neurons & 100 & 100 \\
        HL activation function & ReLU & ReLU \\
        output layer (OL) dimension & 1 & 1 \\
        OL activation function & \textbf{linear} & \textbf{sigmoid} \\
        loss function & \textbf{mean squared error} & \textbf{binary cross entropy} \\
        optimizer & Adam, $\beta_1 = 0.9$, $\beta_2 = 0.999$ & Adam, $\beta_1 = 0.9$, $\beta_2 = 0.999$ \\
        learning rate & 0.001 & 0.001 \\
        epochs & 1000 & 1000 \\
        \bottomrule
    \end{tabular}
    \caption{\label{tab:2HDMscan}%
        Hyperparameters used for sampling the 2HDM-II with regressor and classifier.
        The parts that differ between regressor and classifier are shown in boldface text.
        }
\end{table}

Since the ML classifier is only concerned with points passing the conditions mentioned above,
it is less affected by a fast changing total $\chi^2$
than the regressor

Therefore, we can consider wider ranges without a dramatic increase in the required initial number of points.
Considering this, we scan over the following ranges
\begin{equation}
\begin{gathered}
    0\le\lambda_1\le 10,\qquad
  0\le\lambda_2\le 0.2,\qquad
  -10\le\lambda_3\le 10,\qquad
  -10\le\lambda_4\le 10,\\
    -10\le\lambda_5\le 10,\qquad
  5\le\tan\beta\le 45,\qquad
    -3000 \text{ GeV}^2\le m^2_{12}\le 0 \text{ GeV}^2\,,
\end{gathered}
\end{equation}
where positive $\lambda_1$ and $\lambda_2$ are required for vacuum stability.
In order to keep the light Higgs, $h_1$, as the SM-like Higgs of the model we consider a rather narrow range for $\lambda_2$ and use negative values for the soft $Z_2$-breaking mass parameter $m^2_{12}$.

We use a sequential MLP with four hidden layers with 100 neurons in each layer and ReLU activation function. The output layer has one neuron with Sigmoid activation which maps the output to probabilities between 0 and 1. We train during 1000 epochs. The loss function, binary cross entropy, is minimized by Adam optimizer with learning rate of 0.001 and exponential decay rates $\beta_1 = 0.9$ and $\beta_2 = 0.999$. In each iteration we select $K$ points from the ML predictions. The sampled $K$ points are then passed to the HEP packages to classify them. In the steps where the data sets are imbalanced, SMOTE is automatically called to oversample the minority class.
Additionally, the training data are normalized according to the standard normal distribution.\footnote{The MLP model is very sensitive to the ranges of the input features and we have to normalize the input before we use it to fit the model. Other models, like random forest, are robust against the outliers and can be used without normalization of the input features.}

For the ML regressor, We use a sequential MLP with four hidden layers each with 100 neurons and ReLU activation function. The final output layer contains only one neuron with linear activation function. The loss function in this case is the mean squared error which is minimized by Adam optimizer with learning rate of 0.001 and exponential decay rates $\beta_1 = 0.9$ and $\beta_2 = 0.999$. We train during 1000 epochs in every step. The collected samples are fully utilized to train the ML model after every two iterations, without validation and test samples, since calculating observables precisely using HEP packages is time consuming. 

As for the toy model, here we compare against sampling using MCMC and MultiNest.
Considering that in this case we do not have several identically shaped regions as in the case of the toy model,
we compare against the efficiency in every iteration, defined as the number of in-target points
over the number of tried points.
For the MCMC we use \texttt{emcee}, we start with 300 in-target points with walkers 300 walkers using as log(likelihood) function the \texttt{HiggsSignals} $\chi^2$ plus all the other constraints discussed above.
For Multinest we use \texttt{PyMultiNest} with log-likelihood function as the MCMC\@.

\begin{figure}[tb]
    \centering
    \includegraphics[scale=0.43]{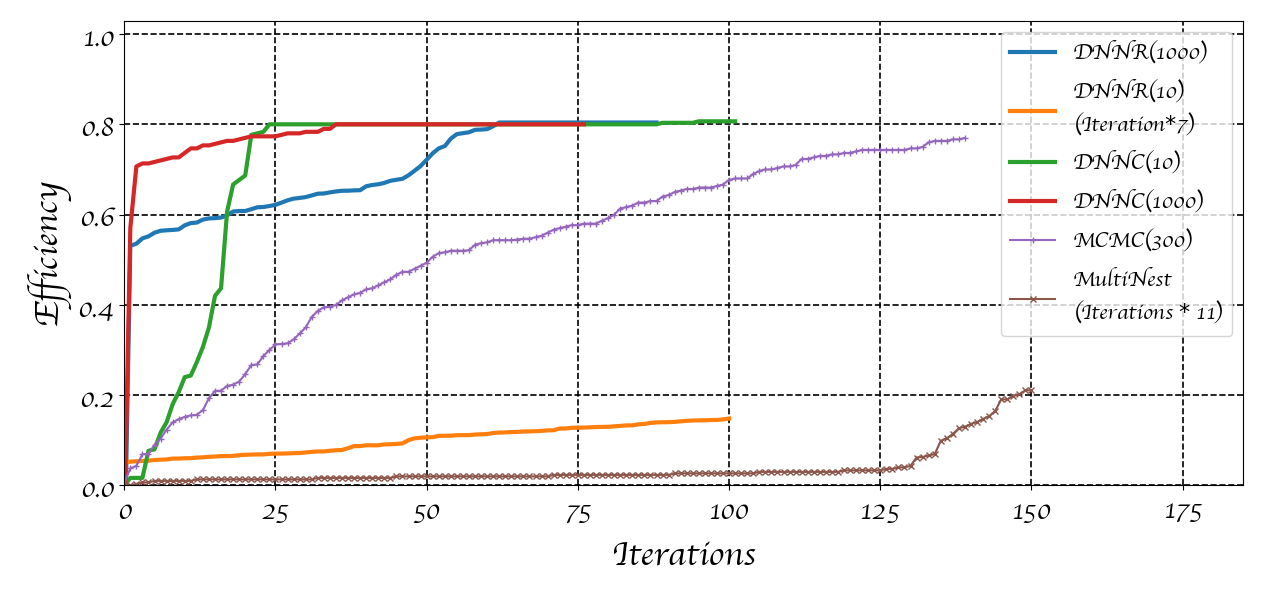}
    \caption{%
        Efficiency in every iteration for different methods.
        For all the methods we run enough iterations to accumulate 20\,000 points.
        The number in parenthesis indicates the initial in-target points for each method, except MultiNest.
        For some methods we have scaled the iterations by a factor indicated as ``Iteration*$x$'',
        meaning actual iterations is number in $x$-axis times this number.
        }
    \label{fig:eff}
\end{figure}

In Fig.~\ref{fig:eff} we show the efficiency for the 4 different methods,
ML classifier (DNNC), regressor (DNNR), MCMC and MultiNest as the number of collected points per iteration over batch size (300). For all methods we require to collect 20\,000  points.

Considering that the convergence efficiency for the DNNC and DNNR depends on the initial number of in-target points, we compare the efficiency with different sizes for initial in-target points, 10 and 1000 points. For both, DNNC and DNNR, we sample the batch $K = 300$  from $L = 200\,000$ with $20\%$ random points. For DNNR with 1000 initial in-target points (blue line) we accumulate 20\,000 points after 88 iterations while for DNNC with the same initial in-target points (red line) it requires 77 iterations.
Here we point out that the maximum efficiency is $\sim 80\%$ since $20\%$ points are chosen randomly in each iteration.
DNNR with 10 initial in-target points (orange line)
has a far slower convergence requiring 700 iterations to accumulate 20\,000 points in-target.
In the case of the DNNC with 10 initial in-target points (green line), the SMOTE technique suggests enough new synthetic points in the target region, resulting in an increase of efficiency.
For this case, efficiency is calculated after correcting with SMOTE (see Fig.~\ref{fig:charts}).
This case requires 100 iterations to accumulate the 20\,000 points.
For MCMC accumulating 20\,000 points requires 140 iterations,
adding each time more in target points, although, at a lower rate than DNNR(1000), DNNC(10) and DNNC(1000).
And MultiNest expectedly spends several iterations in the beginning exploring the regions with
lower likelihood, requiring a total of 1560 iterations.

\subsection{MLP classifier results}

As already discussed,
MCMC and MultiNest require the evaluation
of the log-likelihood from the HEP package
for all the tested points,
which may lead to longer computation time
when acceptance is low.
Moreover, convergence for the DNNR
depends heavily on having a big enough set of in-target points.
This means that for large sampling space
and small target region
the regressor model tends to take longer to start converging.
In the case of the DNNC, we handle this problem with SMOTE
as technique to collect more in-target points in the first steps
and accelerate the initial convergence.
As the DNNC method shows better performance on efficiency as iterations accumulate,
we show the allowed ranges for the 2HDM-II scalar potential parameters
and physical observables from our DNNC scan.

\begin{figure}[tbh!]
    \centering
    \includegraphics[scale=0.54]{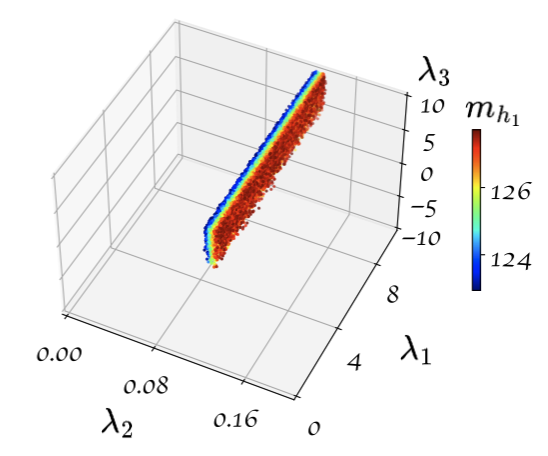}\hspace{1.0cm}
    \includegraphics[scale=0.5]{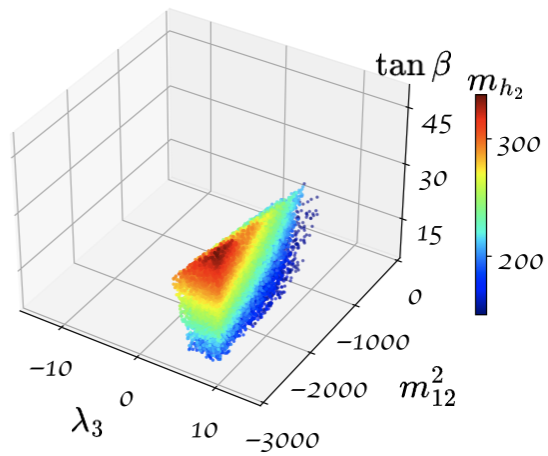}\\
    \vspace{0.5cm}
    \includegraphics[scale=0.5]{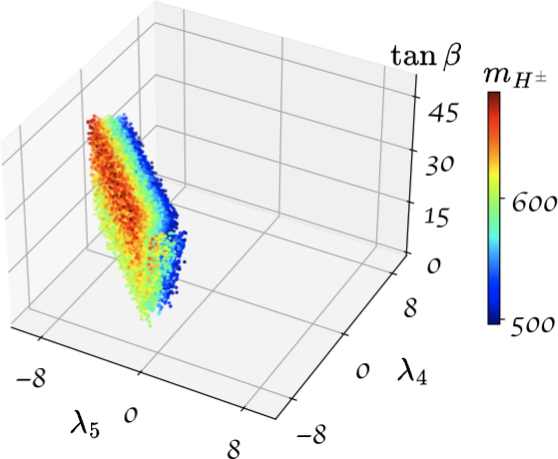}\hspace{1.0cm}
    \includegraphics[scale=0.5]{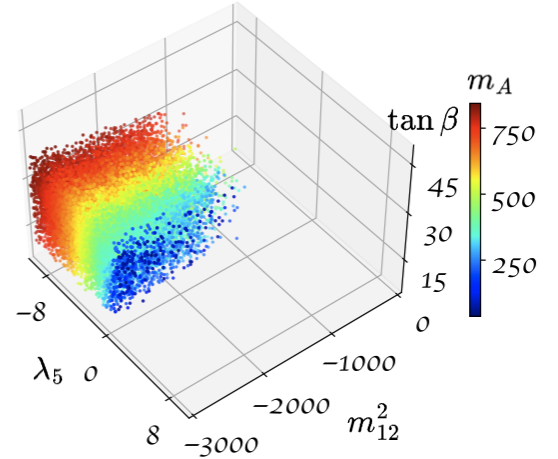}
    \caption{%
        Allowed scalar potential parameters with 20\,000 points.
        Color represents the mass of physical scalars according to the corresponding colorbar.
        All masses are given in GeV ($m^2_{12}$ in GeV$^2$).
        }
\label{fig:ML_Mass}
\end{figure}

Accumulated points satisfy all theoretical and experimental constraints mentioned in Secs.~\ref{sec:themodel} and~\ref{sec:addconst}.
The obtained allowed ranges are the following
\begin{equation}
\begin{gathered}
    1.3\times 10^{-3}\le\lambda^{\text{allowed}}_1\le 10,\quad
  0.12\le\lambda^{\text{allowed}}_2\le 0.14,\quad
  4.1\le\lambda^{\text{allowed}}_3\le 9,\\
  -10\le\lambda^{\text{allowed}}_4\le 5.3,\quad
    -10\le\lambda^{\text{allowed}}_5\le 1.55,\quad
      10\le\tan\beta^{\text{allowed}}\le 38,\\
  -3000 \text{ GeV}^2\le (m^2_{12})^{\text{allowed}}\le -768 \text{ GeV}^2\,.
\end{gathered}
\end{equation}
Considering these ranges for the scalar potential parameters,
in Fig.~\ref{fig:ML_Mass} we show 3-dimensional projections
for several parameters
with color for the masses of different physical scalars.
The SM-like Higgs, $h_1$, the lightest scalar in our setup,
shows a large dependence on $\lambda_2$,
that results in a narrowly distributed region
in the upper left panel of Fig.~\ref{fig:ML_Mass}.
This sharp dependence in $\lambda_2$ can be explained with Eq.~\eqref{eq:mass22}
where, for large $\tan\beta$, the largest contributions comes from $2 \lambda_2 v^2 \sin^2\beta$.
Considering that large $\tan\beta$ implies $\sin\beta \sim 1$ and $v^2$ is fixed,
from Eq.~\eqref{eq:massh1h2} we can deduce that $m_{h_1}$ depends heavily on the value of $\lambda_2$.
Conversely, the mass of the heavier Higgs, $h_2$,
depends on $m^2_{12}$ and $\tan\beta$ through $\chi^2_{11}$ in Eq.~\eqref{eq:massh1h2},
as shown in the upper right panel of Fig.~\ref{fig:ML_Mass}.
For the charged Higgs, $H^\pm$, the mass depends on a combination of $m^2_{12}$, $\lambda_4$, $\lambda_5$ and $\beta$.
In the lower left panel of Fig.~\ref{fig:ML_Mass} can be seen clearly
that $m_{H^\pm}$ depends on the combination of $\lambda_4$ and $\lambda_5$.
It is also possible to see that larger values of $m_{H^\pm}$ are obtained for larger $\tan\beta$.
In the case of the mass of the pseudoscalar, $A$, it depends on a combination of $\lambda_5$, $m^2_{12}$ and $\beta$.
Considering the scan range for $m^2_{12}$,
the pseudoscalar mass is strongly sensitive to $\lambda_5$
as shown the lower right panel of Fig.~\ref{fig:ML_Mass}.

\begin{figure}[tbh!]
    \centering
    \includegraphics[scale=0.3]{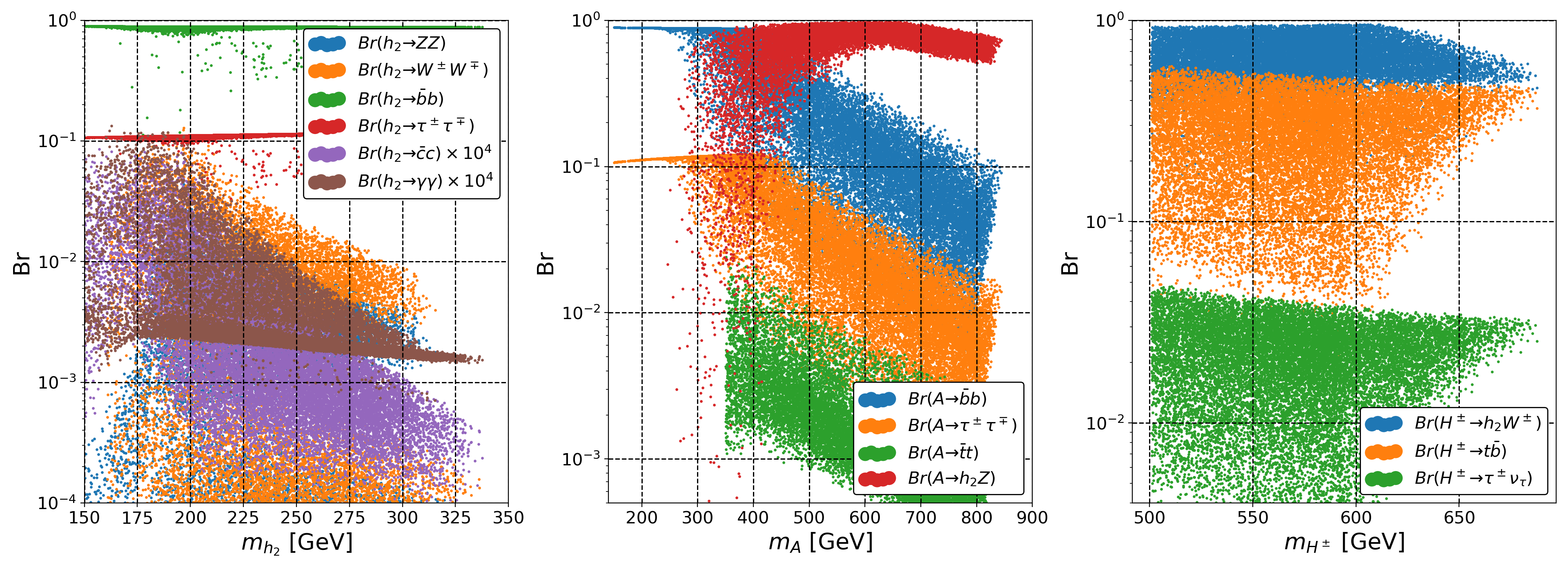}
    \caption{%
        Dependence of branching ratios on the masses of the BSM scalars,
        $h_2$ (left), $A$ (middle) and $H^\pm$ (right),
        using the accumulated 20\,000 points.
        Branching ratios for the SM-like Higgs, $h_1$, are fixed by \texttt{HiggsSignals} to experimental measurements.
    }
\label{fig:THDMC}
\end{figure}

Branching ratios for $h_2$, $A$ and $H^\pm$ scalar bosons are shown in Fig.~\ref{fig:THDMC}.
In the case of the SM-like Higgs, $h_1$, branching ratios have been fixed to experimental measurements by \texttt{HiggsSignals}.
The dominant decay of $h_2$ is via $b$ quark pair and $\tau$ lepton pair,
both proportional to $\cos\alpha/\cos\beta$.
Here, $\alpha$ comes from the combination $\alpha - \beta$,
which is the angle that diagonalizes the CP-even Higgses squared mass matrix.
The di-gauge boson decay of $h_2$ is suppressed by $\cos(\alpha-\beta)$ and hence subdominant,
as shown in the left pane of Fig.~\ref{fig:THDMC}.
This supression is expected,
since it is a consequence of experimental constraints
that force the 2HDM-II closer to the decoupling limit,
$\cos(\alpha - \beta) \sim 0$.
In the middle panel of Fig.~\ref{fig:THDMC},
we show the branching ratios of the pseudoscalar, $A$,
which, for $m_A \lesssim 300$~GeV, is dominated by $b$ quark pair and $\tau$ lepton pair.
This is expected since $A$ decaying to pair of down-type quark pair or lepton pair
is proportional to $\tan\beta$ which we assume to be large.
In the case of decays into pairs of up-type quarks, like $t$, we have suppression by $\cot\beta$.
The dominant decay mode for $m_A \gtrsim 400$~GeV is Br$(A\to h_2 Z)$,
whenever $m_A > m_{h_2} + m_Z$,
which is proportional to $\sin(\alpha-\beta)$.
For the charged Higgs,
decay is dominated by Br$(H^\pm\to h_2 W^\pm)$ which is proportional to $\sin(\alpha-\beta)$,
while the fermionic decays are suppressed by $\cot\beta$,
as can be seen in the right side of Fig.~\ref{fig:THDMC}.

\section{Conclusion}
\label{sec:conclusion}

In this paper we have discussed
the implementation of two broad types of ML based approaches
for efficient sampling of multidimensional parameter spaces:
regression and classification.
The ML model is employed as part of an iterative process
where it is used to suggest new points
and trained according to the results of this suggestion.
In the case of the regression we train a ML model
to precisely predict observables in regions favoured by observations.
For the classification we train the model
to be able to separate the points that belong to the region of interest
from those outside of it.
In the case of the classification
we devise a process to alleviate undersampling of small regions
employing SMOTE\@.
We find that both approaches can efficiently sample regions of interest
with features like several disconnected regions
or regions of low interest inside regions of high interest.
In particular, we applied the two types of model to a 3-dimensional toy model
using sphere-like shell regions as target space.
We compared sampling in this toy model
against results from other methods
used to determine relevant regions in parameter space,
namely MCMC (as implemented in \texttt{emcee})
and MultiNest.
We found that, for both regressor and classifier,
it is possible to achieve a uniform sampling
of all regions of interest with comparable or better distribution.
Moreover, for the classification model we found that using the SMOTE technique
to balance the sampling of the classes
can considerably improve the accumulation of points
when compared to not applying any balancing method.
To illustrate results for a HEP model,
we sampled the parameter space of the 2HDM
by integrating our iterative implementation with popular HEP packages
that take care of calculating the theoretical and experimental results.
We found that we can accumulate points on regions of interest, defined by $\chi^2$
ranges, even when some parameters are narrowly distributed
inside the scanned region.
In particular, the classifier using the SMOTE technique to accelerate
convergence of the model in the first few iterations
can rapidly approach maximum efficiency. 
We compare against the efficiency obtained
when sampling with MCMC and MultiNest.
Expectedly, efficiency for our classifier is much higher
since it is designed to concentrate on the target region.
However, the use of neural networks and iterative training
allow uniform sampling of parameters
regardless of whether their distribution is wide or narrow.
We finalize showing results for the sampled parameter space,
including masses of the physical scalars
and their obtained branching fractions.
Note that there is plenty of space for extensibility,
beyond the characteristics of the employed neural networks.
For example, different types of problems
may benefit from more sophisticated choices of points for training,
using information like disagreement between the trained model
and the full calculation.
This possibilities are left for future improvement
of the techniques described in this work.
The code used for the examples presented here and corresponding documentation is freely available on the web%
\footnote{\url{https://github.com/AHamamd150/MLscanner}}.

\section*{Acknowledgements}
This work is supported by NRF-2021R1A2C4002551. PS is also supported by the Seoul National University of Science and Technology.

\end{document}